\def\thisversion{19 December 2018}
\documentclass[aps,prc,twoside,twocolumn,%
                nofootinbib,showpacs]{revtex4-1}

\usepackage{amsmath,amssymb}
\usepackage{graphicx}

\graphicspath{{figs/}}

\newcommand*{\fs}[1]{#1\!\!\!/}
\newcommand*{\qtext}[1]{\quad\text{#1}\quad}
\newcommand*{\Tpi}{T_\pi}

\newcommand*{\gt}{\gamma_{\textsc{t}}}
\newcommand*{\st}{\sigma_{\textsc{t}}}
\newcommand*{\GDm}[1]{ \left\{ #1 \right\}^\mu }
\newcommand*{\GDn}[1]{ \left\{ #1 \right\}^\nu }
\newcommand*{\up}[1][p]{\underline{#1}}

\newcommand*{\toff}{{\text{off}}}
\newcommand*\qbox{\quad\mbox{}}

\newcommand*{\hA}[1][A]{\hat{#1}}
\newcommand*{\pinch}[1][1]{\mspace{#1mu}}
\newcommand*{\ppinch}{\mspace{2mu}}

\allowdisplaybreaks

\begin{document}

\title{Electromagnetic currents for dressed hadrons}

\author{Helmut~Haberzettl}

 \email{helmut.haberzettl@gwu.edu}

 \affiliation{Institute for Nuclear Studies and Department of Physics, The
 George Washington University, Washington, DC 20052, USA}

\date{\thisversion}

\begin{abstract}
We propose an extension of the minimal-substitution prescription for coupling
the electromagnetic field to hadronic systems with internal structure. The
resulting rules of extended substitution necessarily distinguish between
couplings to scalar and Dirac particles. Moreover, they allow for the
incorporation of electromagnetic form factors for virtual photons in an
effective phenomenological framework. Applied to pions and nucleons, assumed
to be fully dressed to all orders, the resulting dressed currents  are shown
to be locally gauge invariant. Moreover, half-on-shell expressions of (hadron
propagator)$\times$(electromagnetic current) needed in all descriptions of
physical processes will lose \emph{all} information about hadronic dressing
for real photons. The Ball-Chiu ansatz for the spin-1/2 current is seen to
suffer from an incomplete coupling procedure where some essential aspects of
the Dirac particle are effectively treated as those of a scalar particle.
Applied to real Compton scattering on pions and nucleons, we find that
\emph{all} dressing information cancels exactly when external hadrons are on
shell, leaving only gauge-invariant bare Born-type contributions with
physical masses. Hence, nontrivial descriptions necessarily require
contact-type two-photon processes obtained by hadrons looping around two
photon insertion points.
%
\end{abstract}



\maketitle

\section{Introduction}  \label{sec:introduction}

The photon arguably provides the cleanest probe of hadronic structure available
in experiments. As reviewed in Ref.~\cite{KR2010}, real and virtual photon
probes have been and are being used in many experimental facilities around the
world --- JLab, MAMI, ELSA, SPring-8, GRAAL, and others --- to help unravel the
internal dynamics of hadronic systems.

On the theoretical side, however, the situation is much less clean because of
the effective nature of the hadronic degrees of freedom that appear in
experiments. While the electromagnetic interaction is understood perfectly well
at the elementary level, its application to composite baryonic and mesonic
systems of elementary particles is not straightforward. Notwithstanding these
problems, elaborate and sophisticated expansion and power-counting schemes have
been devised to permit the extraction of meaningful model-independent results
from the experimental data (see
Ref.~\cite{Man2018,Petrov:2016azi,Scherer:2012xha} and references therein).
However, given the nature of such schemes, while they work well at low
energies, their application becomes increasingly difficult away from threshold.
At intermediate energies, in particular, where most baryonic states with
nontrivial structure are found, one oftentimes must resort to effective
Lagrangian formulations because they provide a more direct access to the actual
hadronic degrees of freedom --- mesons and baryons --- as they manifest
themselves in experiments.

The present work addresses the question of how to implement the electromagnetic
interaction in an effective Lagrangian formulation where the descriptive
degrees of freedom are mesons and baryons, however, assuming that they are
fully dressed.

Minimal substitution is the standard way of coupling the electromagnetic field
$A^\mu$ to a charged particle with momentum $p$. The corresponding replacement
rule,
\begin{equation}
  p^\mu \to p^\mu -Q A^\mu~,
  \label{eq:MSdefined}
\end{equation}
where $Q$ is the particle's charge operator, is based on the usual covariant
derivative of quantum field theory~\cite{Wtext,PStext}. One of the especially
attractive features of minimal substitution is that it will preserve local
gauge invariance as a matter of course if implemented consistently. However,
while well defined in limited circumstances, it cannot be relied upon to
produce correct or consistent results in all situations~\cite{JMT2013}. In
effective theories, in particular, that subsume elementary QCD degrees of
freedom in terms of hadronic ones, many, if not most, of the shortcomings of
minimal substitution can be traced back to the incorrect or incomplete
implementation of magnetic and polarization properties that arise from the
interaction of the electromagnetic field with the extended charge structure of
the hadrons. The coupling operators for such interactions are transverse and
therefore gauge invariance is not affected by such incomplete implementations,
but the correct description of the physics at hand may suffer.

In the minimal-substitution framework, the currents $J^\mu$ describing the
first-order interaction of a system with an external electromagnetic field are
defined by implementing the substitution (\ref{eq:MSdefined}) for the
(connected) Green's function of the system, performing an expansion in $A_\mu$,
and taking the functional derivative $\delta/\delta A_\mu$ for $A_\mu=0$, and
then truncating all external single-particle propagators according to the LSZ
prescription~\cite{LSZ,Wtext,PStext}, which isolates the current $J^\mu$ as the
coefficient operator of the first-order interaction term $J^\mu A_\mu$. It
should be obvious that this procedure cannot be expected to correctly produce
anomalous magnetic moments because those may be present even if the system as a
whole is uncharged. As a case in point, if one considers the neutron, for
example, as a single effective hadronic system, the corresponding current
vanishes and there is no anomalous contribution at all. To produce anomalous
contributions and polarization effects, one needs to explicitly consider the
substructure of the nucleon in terms of meson-loop dressings and explicitly
incorporate all possible interaction terms in a gauge-invariant
manner~\cite{Man2018}. This can be largely understood as an implementation of
minimal substitution at a more microscopic level.

In general, however, minimal substitution seems to be incapable of producing
electromagnetic form factors for virtual photons directly. As has been pointed
out~\cite{KPS}, the dressing effects due to minimal substitution --- even when
applied to fully dressed hadronic entities --- will depend on the squared
hadronic four-momenta going into and coming out of the electromagnetic vertex,
but they cannot produce the clean $k^2$ dependence required to produce the
electromagnetic form factors $F(k^2)$, where $k$ is the photon four-momentum.
Hence, the $k^2$ dependence of form factors --- usually simply stated as a
requirement based on Lorentz invariance --- must result from other dynamical
effects. Rather than discussing the possible nature of such effects, we point
out that we cannot expect to be able to describe form-factor effects in any
standard application of minimal substitution in an effective dynamical
framework.

Given this situation, we propose here to turn the question around and ask how
one can extend minimal substitution to incorporate known experimental
information --- masses, charges, anomalous moments, and $k^2$ dependence of
form factors --- into a framework that assumes that \emph{all} of the hadronic
dressing effects of system are known to all orders. How such dressing effects
are obtained is then a problem secondary to constructing current operators that
are consistent with the assumed hadronic dressing.

We will show that this consistency requirement means that the proposed
\textit{extended substitution} must distinguish in essential aspects between
how the field couples to scalar (spin-0) particles and Dirac (spin-1/2)
particles. We will construct the fully dressed electromagnetic currents for
pions and for nucleons within the proposed framework which will incorporate
known experimental information about these hadrons. For the nucleon current,
this will remedy a shortcoming of the spin-1/2 Ball-Chiu ansatz~\cite{BC80}.
Specifically, we will show that the Ball-Chiu current results from treating
some essential aspects of the nucleon dressing as being that of a scalar
particle, which clearly is inconsistent.

We will also show that products of (hadron propagator)$\times$(electromagnetic
current) --- both fully dressed --- will lose \emph{all} information about
their detailed dressing mechanisms when taken half-on-shell on the current
side, with the \emph{physical} hadron mass being the only remnant of the
dressing and  additional off-shell information only entering for virtual
photons. For meson photoproduction processes with real photon, in particular,
this means that for the usual $s$-, $u$-, and $t$-channel Born terms depicted
in Fig.~\ref{fig:photo} only the meson-production vertex carries any structure
information. Complete cancellations of off-shell effects also extend to the
respective real Compton scattering reactions for both pions and nucleons,
leaving only Born-type expressions. It is argued that nontrivial
Compton-scattering contributions require genuine two-photon processes where the
respective (off-shell) Compton tensor is dressed by hadron loops.

The paper is organized as follows. In Sec.~\ref{sec:extsub}, we will present
the basic rules for the proposed extension of minimal substitution. We will do
so using the device of the gauge derivative of Ref.~\cite{hh97} which provides
a convenient shorthand notation for the implementation of minimal substitution.
(For completeness, some pertinent details of the gauge-derivative procedure are
recapitulated in the Appendix.) In Secs.~\ref{sec:pion} and \ref{sec:nucleon},
respectively, we apply the proposed extended substitution rules to the fully
dressed pion and nucleon propagators as examples for spin-0 and spin-1/2
particles and construct their corresponding fully dressed current operators.
For the nucleon case, we show that the Ball-Chiu current ansatz~\cite{BC80}
suffers from an incomplete coupling procedure. We also derive the
aforementioned cancellation of dressing effects for half-on-shell combinations
of propagator and current for real photons, which extends to the on-shell
Compton tensors. The final Sec.~\ref{sec:summary} provides a summary and
discussion of the present findings.

\section{Rules of Extended Substitution}\label{sec:extsub}

Following Ref.~\cite{hh97}, we will use the device of the gauge derivative as a
shorthand tool for describing how minimal substitution affects the reaction
dynamics of a particle with momentum $p$ and associated charge operator $Q$. As
the examples in Refs.~\cite{hh97,HNO18} demonstrate, the gauge derivative may
be used to consistently link all topological elements of reaction mechanisms
that contribute to the interaction with the external electromagnetic field, in
a procedure sometimes referred to as `gauging of equations'. However, for the
present purpose, we only need to consider the `last step', when the gauge
derivative is applied to the functions
--- propagators, vertices --- that describe elements of the reaction at hand to
provide the actual coupling mechanisms $J^\mu$ to the external field $A_\mu$.

With the rules given in Ref.~\cite{hh97}, briefly summarized here in
Appendix~\ref{app:MS} for completeness, for a spin-0 scalar particle of
momentum $p$ and charge $Q$, the current operator results from
\begin{equation}
  \GDm{p^2} = Q(p'+p)^\mu~,
  \label{eq:MSbasicPion}
\end{equation}
where $p'=p+k$ for an (incoming) photon with four-momentum $k$. The
gauge-derivative braces $\GDm{\cdots}$ here indicate coupling of the photon
four-momentum $k^\mu$ to the functional dependence $p^2$. The result
(\ref{eq:MSbasicPion}) immediately follows from Eq.~(\ref{eq:GDpropagator}) in
the Appendix and the fact that the inverse propagator of a scalar particle is a
function of $p^2$ alone. For a spin-1/2 Dirac particle, we also find the usual
coupling mechanism
\begin{equation}
  \GDm{\fs{p}}=Q\gamma^\mu
  \label{eq:MSbasicNucleon}
\end{equation}
because its inverse propagator is a function of $\fs{p}$.

Considering now the gauge derivative of an invariant scalar functions $f(p^2)$
of the particle's squared four-momentum, clearly, we have
\begin{align}
  \GDm{f(p^2)}_S &= \GDm{p^2} \frac{f(p'^2)-f(p^2)}{p'^2-p^2}
  \nonumber\\
  &=Q(p'+p)^\mu \frac{f(p'^2)-f(p^2)}{p'^2-p^2}~,
  \label{eq:MSscalar0}
\end{align}
as expected, where the index $S$ indicates that $f(p^2)$ results from the
dynamics of a scalar particle. The proof is easily found by expanding $f(p^2)$
in powers of $p^2$, applying the product rule to every term in the expansion,
and then resumming. (This assumes that the expansion is well-defined at least
at the formal level. In general, nonanalytic functions may require special
treatments.) The function ratio on the right-hand side of
Eq.~(\ref{eq:MSscalar0}) presents a well-defined $\frac{0}{0}$ situation at
$p'^2=p^2$ providing the derivative of $f$. Because of this result, such
finite-difference derivatives (FDDs) will be ubiquitous in the present
investigation.

For a Dirac particle, we may write $\fs{p}^2 = p^2$, which is crucial to
expressing the corresponding Feynman propagator as
\begin{equation}
  \frac{1}{\fs{p}-m} = \frac{\fs{p}+m}{p^2-m^2}
  \label{eq:DiracKG}
\end{equation}
to establish that, in addition to the Dirac equation, it also solves the
Klein-Gordon equation for the same mass $m$. One finds this equivalence would
be destroyed if one applied the scalar gauge-derivative result
(\ref{eq:MSbasicPion}) to the $p^2$ dependence on the right-hand side of this
equation. Instead, as explained in the context of Eq.~(\ref{eq:demandequality})
in the Appendix, equivalence demands that with
\begin{align}
\GDm{\fs{p}^2}&= \fs{p}'\GDm{\fs{p}}+\GDm{\fs{p}}\fs{p}
\nonumber\\
&=Q(\fs{p}' \gamma^\mu + \gamma^\mu\fs{p})~,
\end{align}
which follows from the  product rule (\ref{eq:GDbasicproduct}), one needs to
introduce the Dirac version
\begin{equation}
  \GDm{f(p^2)}_D = \GDm{\fs{p}^2} \frac{f(p'^2)-f(p^2)}{p'^2-p^2}
  \label{eq:GDDirac}
\end{equation}
to replace Eq.~(\ref{eq:MSscalar0}) for Dirac particles. The relationship
between $\GDm{f}_D$ and $\GDm{f}_S$ is given by the Gordon identity,
\begin{equation}
\fs{p}' \gamma^\mu + \gamma^\mu\fs{p}  = (p'+p)^\mu +i\sigma^{\mu\nu}k_\nu~,
\label{eq:Gordon}
\end{equation}
i.e., the two derivatives differ by a manifestly transverse term. Their
respective four-divergences,
\begin{equation}
  k_\mu \GDm{f(p^2)}_S = k_\mu \GDm{f(p^2)}_D = Q\left[ f(p'^2)-f(p^2)\right] ~,
\end{equation}
therefore, are unaffected which is crucial for maintaining gauge invariance.

\subsection{Extending minimal substitution}

The basic coupling mechanisms for scalar and Dirac particles described so far,
if implemented consistently, are sufficient to provide a current for any system
that maintains local gauge invariance as expressed in terms of Ward-Takahashi
identities (WTI) for three-point functions~\cite{WTI} and generalized WTIs for
$N$-point functions~\cite{Kazes,hh97,HNO18}. However, as discussed in the
Introduction, they cannot account for any structure that results from
electromagnetic form factors. Such effects, therefore, must come from
manifestly transverse coupling mechanisms.

For definiteness, we will here consider the pion as an example of a scalar
particle, identified by index $\pi$, and the nucleon as a spin-1/2 Dirac
particle, with two charge states $N=p,n$, where
\begin{equation}
  Q_p = e\frac{1+\tau_3}{2} \qtext{and} Q_n = e\frac{1-\tau_3}{2}
  \label{eq:Nisospin}
\end{equation}
are the respective isospin operators for proton ($p$) and neutron ($n$),
respectively, with $\tau_3$ being the usual Pauli matrix; $e$ is the
fundamental charge unit.

For the pion with four-momentum $q$, we amend the elementary scalar coupling
rule (\ref{eq:MSbasicPion}) and allow for
\begin{equation}
  \GDm{q^2} = Q_\pi (q'+q)^\mu + \Tpi^\mu(q',q)~,
  \qtext{with} k_\mu \Tpi^\mu \equiv 0~,
  \label{eq:MSPionextended}
\end{equation}
where the transverse current is given by
\begin{equation}
\Tpi^\mu(q',q) = Q_\pi t^\mu(q',q) f_\pi(q',q)~,
\label{eq:Tpitransverse}
\end{equation}
where
\begin{equation}
  t^\mu(q',q) = (q'+q)^\mu -k^\mu\frac{q'^2-q^2}{k^2}
  \label{eq:transversescalar}
\end{equation}
is the only manifestly transverse operator one can construct with two
independent four-momenta $q$ and  $q'=k+q$.  The scalar (and symmetric)
form-factor function $f_\pi(q',q)=f_\pi(q'^2,q^2;k^2)$ here must vanish at
$k^2=0$ to make the current nonsingular. More details about its relationship to
the physical (on-shell) pion form factor $F_\pi(k^2)$ will be given in the
subsequent Sec.~\ref{sec:pion}. $Q_\pi$ here describes the charges of the pions
in units of $e$, with $f_\pi$ being the same for $\pi^\pm$; $\pi^0$ has no form
factor because it is its own antiparticle.

For the nucleon $N$ with four-momentum $p$ and mass $m$, we amend the basic
Dirac-particle rule (\ref{eq:MSbasicNucleon}) by
\begin{equation}
  \GDm{\fs{p}} = \Gamma_N^\mu(p',p)\equiv Q_p \gamma^\mu + T_N^\mu(p',p)~,
  \label{eq:MSNextended}
\end{equation}
where $T_N^\mu$ is a manifestly transverse current,
\begin{equation}
  k_\mu T_N^\mu \equiv 0~,\qtext{for} N=p,n~,
\end{equation}
that can be expressed in terms of the two transverse operators
\begin{equation}
  \st^\mu = \frac{i\sigma^{\mu\nu}k_\nu}{2m}
  \qtext{and}
  \gt^\mu = \gamma^\mu -k^\mu \frac{\fs{p}'-\fs{p}}{k^2}~,
\label{eq:TNtransverse}
\end{equation}
where the latter will require a coefficient function that vanishes at $k^2=0$
to render the corresponding current well defined. The transverse currents then
may be written as
\begin{equation}
T_N^\mu(p',p) = Q_N\left(\gt^\mu f^N_1+\st^\mu f^N_2\right)~, \qtext{for} N=p,n~,
\label{eq:emFFexpand}
\end{equation}
where the four scalar (and symmetric) coefficient functions
$f^N_i=f^N_i(p'^2,p^2;k^2)$ ($N=p,n$; $i=1,2$) are to be constrained by the
Dirac and Pauli form factors of the proton and neutron. More details will be
discussed in the nucleon section~\ref{sec:nucleon} below.

Note here that the $k^\mu$ contributions in (\ref{eq:Tpitransverse}) and
(\ref{eq:TNtransverse}) are necessary for formal reasons to verify the
respective transversality conditions. For practical purposes, however, we may
drop such terms from any physically relevant current since $\epsilon_\mu k^\mu
=0$ for covariant physical photon polarization $\epsilon_\mu$, irrespective of
whether the photon is real or virtual. Moreover, in view of the numerator
expressions $q'^2-q^2$ and $\fs{p}'-\fs{p}$ in the respective transverse
couplings, these terms do not contribute anyway for on-shell hadrons.
Nevertheless, one might be well advised to drop these terms only at the very
end, when physical matrix elements are to be calculated because doing so at the
very start may lead to erroneous conclusions.

To see the effect of the extended substitution rules (\ref{eq:MSPionextended})
and (\ref{eq:MSNextended}) over the respective basic rules
(\ref{eq:MSbasicPion}) and (\ref{eq:MSbasicNucleon}) in the following
considerations, one only needs to put the respective current $\Tpi^\mu$ and
$T_N^\mu$ to zero. Obviously, since the extensions only add transverse
currents, they have no effect on gauge invariance at all.

\section{Pion: Spin 0}\label{sec:pion}

The most general propagator for a fully dressed pion with four-momentum $q$ and
mass $\mu$ can be written as
\begin{equation}
  \Delta_\pi(q^2) = \frac{1}{(q^2-\mu^2)\Pi(q^2)}~,
  \label{eq:PionPropDressed}
\end{equation}
where the scalar dressing function $\Pi(q^2)$ is normalized as
\begin{equation}
\Pi(\mu^2) =1~,
\end{equation}
which provides the required unit residue for the propagator.

%
\begin{figure*}[!t]\centering
  \includegraphics[width = .7\textwidth,clip=]{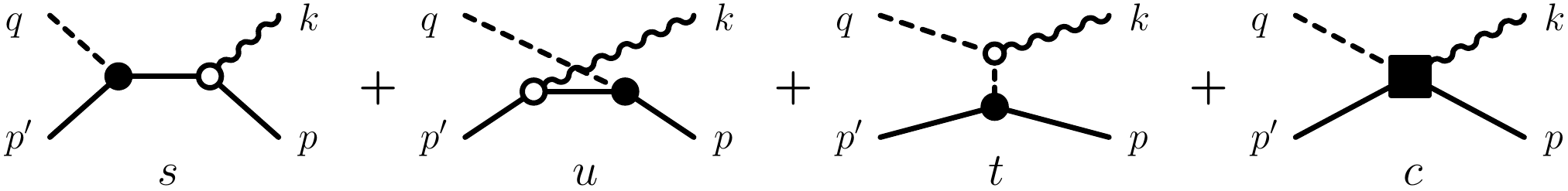}
  \caption{\label{fig:photo}%
  Basic topology of pion photoproduction off the nucleon, with associated
  four-momenta.  Here, and in all other diagrams, time runs from right to left.
  The first three diagrams depict $s$-, $u$-, and $t$-channel, respectively,
  according to the respective Mandelstam variable of the intermediate off-shell
  hadron. The last diagram comprises all nonpolar contact-type mechanisms
  including final-state interactions~\cite{hh97}. The half-on-shell
  contributions for the pion current in the $t$-channel and for the nucleon
  currents in the $s$- and $u$-channels are discussed in
  Secs.~\ref{sec:halfpion} and \ref{sec:halfnucleon}, respectively.}
\end{figure*}
%

We may expand the dressing functions around the pole by writing
\begin{equation}
  \Pi(q^2)   = 1 + \frac{q^2-\mu^2}{\mu^2 }D_\pi(q^2)
  \label{eq:PIexpand}
\end{equation}
where
\begin{equation}
  D_\pi(q^2) = \mu^2 \frac{\Pi(q^2)-1}{q^2-\mu^2}
  \label{eq:FDDpi1}
\end{equation}
is a well-defined nonsingular finite-difference derivative (FDD) that is
proportional to the derivative of the dressing function in the limit $q^2\to
\mu^2$ at the pole. (The proportionality factor $\mu^2$ is only introduced to
make $D_\pi$ dimensionless.) We thus have
\begin{equation}
  \Delta_\pi(q^2) = \frac{1}{q^2-\mu^2} - \frac{D_\pi(q^2)}{\Pi(q^2) \mu^2 }~,
  \label{eq:pionprop0}
\end{equation}
where the only remnant of the dressing in the pole term is the physical mass
$\mu$; all other dressing effects sit in the nonpolar contact-type counterterm.

\subsection{Pion current}

Using the extended substitution (\ref{eq:MSPionextended}), the pion current for
an incoming photon with momentum $k=q'-q$ is obtained from gauging the inverse
propagator according to
\begin{align}
J_\pi^\mu(q',q)
& =\GDm{\Delta_\pi^{-1}(q^2)}_S
\nonumber\\[2ex]
&=\GDm{\frac{q^2 \Pi(q^2) + \Pi(q^2) q^2}{2} -\mu^2 \Pi(q^2)}_S~,
\label{eq:JpionGD}
\end{align}
where the scalar-particle gauge derivative is applied to the symmetrized
expressions as explained in Appendix~\ref{app:MS}. With the scalar
gauge-derivative rule (\ref{eq:MSscalar0}), using the extended substitution
(\ref{eq:MSPionextended}), we then obtain the fully dressed current as
\begin{align}
J_\pi^\mu(q',q)  = \Big[ Q_\pi (q'+q)^\mu +T^\mu_\pi\Big] \frac{\Delta^{-1}_\pi(q'^2)-\Delta_\pi^{-1}(q^2)}{q'^2-q^2}~.
\label{eq:Jpion1}
\end{align}
For $T^\mu_\pi\equiv 0$, this result was given in Ref.~\cite{KPS}. This current
trivially satisfies the appropriate WTI for the pion,
\begin{equation}
  k_\mu J_\pi^\mu(q',q) = Q_\pi \Big[\Delta^{-1}_\pi(q'^2)-\Delta_\pi^{-1}(q^2)\Big]~,
  \label{eq:WTIpion}
\end{equation}
and thus is manifestly locally gauge invariant. We emphasize that this finding
does not result from a condition that needs to be imposed on the current, but
that it is a straightforward consequence of the construction procedure in terms
of the gauge derivative (\ref{eq:JpionGD}). Note that the WTI does not involve
any form-factor information even if $T^\mu_\pi\neq 0$.

We emphasize that the current (\ref{eq:Jpion1}) is constructed here to be
consistent with the propagator (\ref{eq:PionPropDressed}) such that the WTI
follows as a matter of course. The procedure, however, makes no assumption
about the details of the dressing function $\Pi(q^2)$ other than stipulating
that it produces a unit residue. Therefore, any (non-pathological)
determination of $\Pi(q^2)$, whether sophisticated or not, that produces
$\Pi(\mu^2)=1$ will fit the present framework.

With nonzero transverse contributions $T^\mu_\pi$, the current
(\ref{eq:Jpion1}) actually comprises the most general expression one can write
down for the scalar current if one allows for arbitrary \emph{nonsingular}
symmetric expressions for the form factor,
$f_\pi(q',q)=f_\pi(q'^2,q^2;k^2)=f_\pi(q^2,q'^2;k^2)$. Since \emph{any}
additional dressing effect must be transverse and off-shell, they may always be
assumed to be already subsumed in these form factors as a matter of course.
Including on- and off-shell degrees of freedom in these form factors,
therefore, the current (\ref{eq:Jpion1}) comprises all possibilities. For
further discussion of off-shell freedom, see the summarizing
Sec.~\ref{sec:summary}.

Half on shell, the current reduces to
\begin{equation}
  J^\mu_\pi(q',q) = \Big[ Q_\pi (q'+q)^\mu +T^\mu_\pi(q',q)\Big]\Pi(q_\toff^2)~,
  \label{eq:JpiHalfOnShell}
\end{equation}
where $q^2_\toff$ is either $q'^2$ or $q^2$ depending on which leg is off
shell. The fully on-shell current thus reads
\begin{align}
J_\pi^\mu(\up[q]',\up[q]) &= Q_\pi (\up[q]'+\up[q])^\mu +T^\mu_\pi(\up[q]',\up[q])
\nonumber\\
&= Q_\pi (\up[q]'+\up[q])^\mu \big[1+f_\pi(\mu^2,\mu^2;k^2) \big] ~,
\end{align}
where the underlining indicates on-shell momenta. Hence, in view of the unit
result for the physical form factor $F_\pi(k^2)$ for real photons and to ensure
that the current (\ref{eq:Jpion1}) is nonsingular both on and off shell, we may
write
\begin{equation}
  f_\pi(q'^2,q^2;k^2) =\frac{k^2}{\mu^2}H_\pi(q'^2,q^2;k^2)~,
\end{equation}
where the nonsingular symmetric function $H_\pi$ is defined by this equation,
and determine the physical (on-shell) form factor as
\begin{equation}
  F_\pi(k^2) = 1+\frac{k^2}{\mu^2}H_\pi(\mu^2,\mu^2;k^2)~.
\end{equation}
Hence, the coupling operator of the pion current (\ref{eq:Jpion1}) may be
written as
\begin{align}
Q_\pi (q'+q)^\mu +T^\mu_\pi &=Q_\pi (q'+q)^\mu
\left[1+\frac{k^2}{\mu^2}H_\pi(q'^2,q^2;k^2)\right]
\nonumber\\[1ex]
&\quad\mbox{}
-Q_\pi k^\mu \frac{q'^2-q^2}{\mu^2} H_\pi(q'^2,q^2;k^2)~.
\label{eq:PionCouplingExplicit}
\end{align}

\subsection{Half-on-shell contribution}\label{sec:halfpion}

Let us consider the half-on-shell situation of the current for the outgoing
$t$-channel pion with the off-shell intermediate propagator, as depicted in the
photoproduction diagrams of Fig.~\ref{fig:photo}. Using the kinematics of the
figure, with the outgoing pion on shell at $q^2=\mu^2$ and $t=(q-k)^2$ for the
intermediate pion, we have
\begin{align}
J^\mu_\pi(\up[q],q-k)\Delta_\pi(t)
&=\Big[Q_\pi  (2q-k)^\mu +T^\mu_\pi\Big] \frac{1}{t-\mu^2} ~,
\label{eq:JpionHalfOnshell0}
\end{align}
where the propagator dressing completely cancels. Using
(\ref{eq:PionCouplingExplicit}) and the FDD
\begin{equation}
   D_H(t;k^2) =\mu^2 \frac{H_\pi(\mu^2,t;k^2)-H_\pi(\mu^2,\mu^2;k^2)}{t-\mu^2}~,
\end{equation}
we obtain
\begin{align}
J^\mu_\pi(\up[q],q-k)\Delta_\pi(t)
&=Q_\pi\frac{(2q-k)^\mu}{t-\mu^2} F_\pi(k^2)
\nonumber\\
&\mbox{}\quad
+Q_\pi \frac{(2q-k)^\mu}{\mu^2} \frac{k^2}{\mu^2} D_H(t;k^2)
\nonumber\\
&\mbox{}\quad
+Q_\pi \frac{k^\mu}{\mu^2} H_\pi(\mu^2,t;k^2)~.
\label{eq:JpionHalfOnshell2}
\end{align}
This result is exact for the dressed current (\ref{eq:Jpion1}). All hadronic
dressing effects fully cancel here and we are left with the usual expression
resulting from elementary Feynman rules for the pole term. The contact term
depending on the FDD $D_H(t;k^2)$ contributes only for electroproduction.
Again, the contact-type $k^\mu$ term can be ignored for physical matrix
elements, however, it is necessary to provide the correct four-divergence of
this half-on-shell expression,
\begin{equation}
  k_\mu J^\mu_\pi(\up[q],q-k)\Delta_\pi(t) = -Q_\pi~,
\end{equation}
where the minus sign signifies that the on-shell pion is outgoing. This result
is identical to
\begin{equation}
  k_\mu\ppinch Q_\pi (2q-k)^\mu \frac{1}{t-\mu^2} = -Q_\pi
\end{equation}
for the `bare' situation where the only dressing effect is the physical mass
$\mu$. Without the $k^\mu$ term in (\ref{eq:JpionHalfOnshell2}), this
equivalence cannot be established.

We add here that in the four-point-function context of the $t$-channel graph in
Fig.~\ref{fig:photo}, the cancellation of dressing effects does not extend to
the second vertex; in other words, the purely hadronic $\pi NN$ vertex here
retains the dressing that accounts for the intermediate pion being off-shell.

\subsection{Real Compton scattering on the pion}\label{sec:ComptonPion}

\begin{figure}[!t]\centering
  \includegraphics[width=.90\columnwidth,clip=]{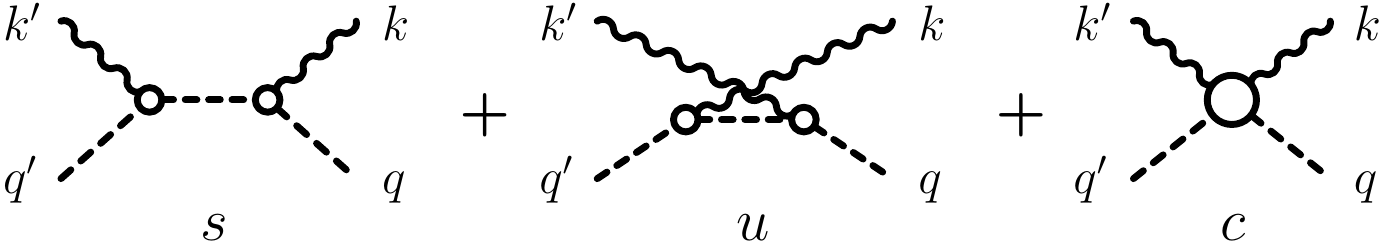}
  \caption{\label{fig:ComptonPion}%
  Diagrams for Compton scattering on a charged pion. In
  Eqs.~(\ref{eq:ComptonPionSU})--(\ref{eq:ComptonPionNoDressing}), the incoming
  and outgoing four-momenta are $k^\mu$ and $k'^\nu$, respectively.}
%
\bigskip
%
  \includegraphics[width=.95\columnwidth,clip=]{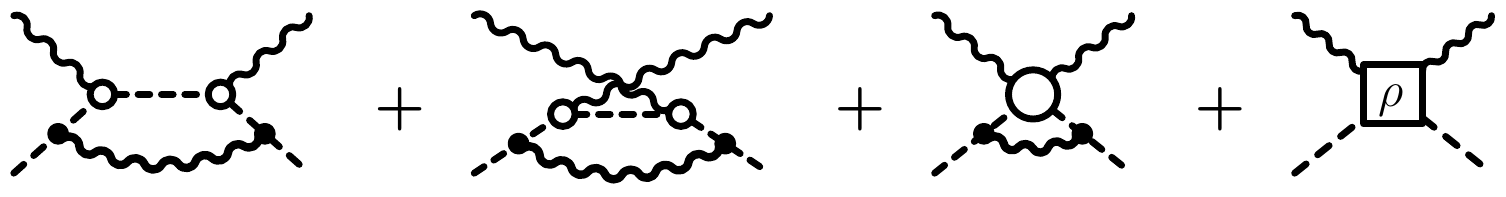}
  \caption{\label{fig:ComptonPionDressed}%
  Example of genuine two-photon contributions to Compton scattering on
  the pion: Crossing-symmetric dressing of the amplitude depicted in
  Fig.~\ref{fig:ComptonPion} by a vector-meson loop. The contact-type
  square labeled $\rho$ in the last diagram subsumes necessary mechanisms that
  make the entire contribution transverse.}
\end{figure}

A particularly straightforward application is given by making the second vertex
an electromagnetic one as well, resulting in Compton scattering on a charged
pion as depicted in Fig.~\ref{fig:ComptonPion}. For real Compton scattering the
situation is particularly simple because there is no electromagnetic form
factor dependence and the only structure is provided by the propagator dressing
function $\Pi(q^2)$. With the half-on-shell result (\ref{eq:JpionHalfOnshell2})
taken for $k^2=0$, dropping the irrelevant $k^\mu$ contribution,  and with
Eq.~(\ref{eq:JpiHalfOnShell}), we easily see that the Compton tensors for the
$s$- and $u$-channel diagrams of Fig.~\ref{fig:ComptonPion}, where $s=(q+k)^2$
and $u=(q-k')^2$ are the Mandelstam variables for the respective intermediate
off-shell particle, read
\begin{subequations}\label{eq:ComptonPionSU}
\begin{align}
M^{\nu\mu}_s(\up[q]',\up[q]) &= e^2 (2q'+k')^\nu\frac{\Pi(s)}{s-\mu^2}(2q+k)^\mu
\nonumber\\
 &= e^2 \frac{(2q'+k')^\nu\, (2q+k)^\mu}{s-\mu^2}
 \nonumber\\
 &\qquad\mbox{} +e^2 (2q'+k')^\nu\frac{D_\pi(s)}{\mu^2}(2q+k)^\mu
 ~,
\\
M^{\nu\mu}_u(\up[q]',\up[q]) &=e^2 (2q'-k)^\mu \frac{\Pi(u)}{u-\mu^2}(2q-k')^\nu
\nonumber\\
&=e^2 \frac{(2q'-k)^\mu \,(2q-k')^\nu}{u-\mu^2}
\nonumber\\
&\qquad\mbox{}
+e^2 (2q'-k)^\mu \frac{D_\pi(u)}{\mu^2}(2q-k')^\nu~,
\end{align}
\end{subequations}
where the respective second equalities provide the decomposition into pure
(undressed) pole contributions and contact terms depending on the FDDs of the
propagator dressing function according to Eq.~(\ref{eq:FDDpi1}). The squared
charge operator $Q_\pi^2$ for $\pi^\pm$ is simply written here as $e^2$.

The contact term in Fig.~\ref{fig:ComptonPion} can be obtained explicitly by
coupling a second photon into the current (\ref{eq:Jpion1}). Rather than doing
this fully off-shell, it is easier to employ the half-on-shell expressions
(\ref{eq:JpiHalfOnShell}), and then symmetrize and apply the gauge derivative
to entities depending on the off-shell momentum $q_\toff$. With an overall
minus sign resulting from the relationship of the gauge derivative to the
functional derivative, as seen from Eq.~(\ref{eq:GDBasicDefinition}), one finds
\begin{align}
  M^{\nu\mu}_c(\up[q]',\up[q]) &= -\frac{1}{2}\left[\GDn{J_\pi^\mu(q_\toff,\up[q])}+\GDm{J_\pi^\nu(q_\toff,\up[q]}\right]
  \nonumber\\
  &\qquad\mbox{}  -\frac{1}{2}\left[\GDn{J_\pi^\mu(\up[q]',q_\toff)}+\GDm{J_\pi^\nu(\up[q]',q_\toff}\right]
  \nonumber\\
  &= -2e^2 g^{\mu\nu}
 -e^2 (2q'+k')^\nu\frac{D_\pi(s)}{\mu^2}(2q+k)^\mu
\nonumber\\
&\qquad\mbox{}
-e^2 (2q'-k)^\mu \frac{D_\pi(u)}{\mu^2}(2q-k')^\nu~.
\label{eq:ComptonPionContact}
\end{align}
The last two terms here cancel the FDD contributions in the $s$- and
$u$-channel terms in (\ref{eq:ComptonPionSU}). Hence, summing up all three
contributions,  the entire Compton tensor for all diagrams in
Fig.~\ref{fig:ComptonPion} for real photons scattering off the pion is given as
\begin{align}
M^{\nu\mu}_\pi(\up[q]',\up[q]) &= e^2 \frac{(2q'+k')^\nu\, (2q+k)^\mu}{s-\mu^2}
\nonumber\\
&\quad\mbox{}
+e^2 \frac{(2q'-k)^\mu \,(2q-k')^\nu}{u-\mu^2}-2 e^2  g^{\mu\nu}
~,
\label{eq:ComptonPionNoDressing}
\end{align}
where the only remnant of the dressing structure is the physical pion mass. All
other effects completely cancel and what remains is just what is usually
referred to as the Born amplitude for real Compton scattering on the pion, with
undressed Feynman-type $s$- and $u$-channel pole terms and the $-2g^{\mu\nu}$
coupling for the contact term. This on-shell tensor is trivially gauge
invariant because of four-momentum conservation, $q'+k'-q-k=0$.

%
%

The present finding of all dressing effects canceling here seems to corroborate
the conjecture of Kaloshin~\cite{Kal99} that this would be the case. This was
disputed in Ref.~\cite{KPS}. Kaloshin's conjecture was based on the limited
evidence of an $s$-wave one-loop calculation with intermediate $\pi\sigma$
states. The derivation of Eq.~(\ref{eq:ComptonPionNoDressing}) shows that this
is indeed true at any order, for any dressing mechanism of the pion. However,
the present construction of the Compton tensor is limited to the part of
Compton scattering that can be understood as two sequential one-photon
processes. Genuine two-photon processes like, for example, the vector-meson
dressing mechanism depicted in Fig.~\ref{fig:ComptonPionDressed} are not taken
into account. Contact-type two-photon currents of this kind, therefore, can be
expected to resolve the discrepancy between Refs.~\cite{Kal99} and \cite{KPS}.

We emphasize that the complete cancellation of all dressing effects found here
is true only for real photons and when the external pions are on shell. The
latter requirement is not true for the dressed Compton tensor in
Fig.~\ref{fig:ComptonPionDressed}, which therefore retains its off-shell
structure. We add that on-shell cancellations are not particular to the pion;
in Sec.~\ref{sec:ComptonNucleon} below, we will find a similar result for real
Compton scattering on the nucleon.

\section{Nucleon: Spin 1/2}\label{sec:nucleon}

Without lack of generality, the dressed spin-1/2 propagator for the nucleon
with physical mass $m$ and four-momentum $p$ may be written as
\begin{equation}
S(p) = \frac{1}{\fs{p}A(p^2)-mB(p^2)}~.
\label{eq:NucProp}
\end{equation}
The two independent scalar dressing functions $A$ and $B$ are constrained by
\begin{equation}
A(m^2)=B(m^2)~,
\label{eq:A=B}
\end{equation}
which ensures the propagator has a pole at $\fs{p}\to m$, and
\begin{equation}
A(m^2)+2m^2\frac{d\left[A(p^2)-B(p^2)\right]}{dp^2}\bigg|_{p^2=m^2}=1
 ~,
 \label{eq:residue}
\end{equation}
which provides a unit residue for this pole.

Even though we will not make use of it here, we note that, without lack of
generality, we may write
\begin{subequations}\label{eq:APparam}
\begin{align}
A(p^2) &= \frac{p^2+(2a-1)m^2}{2m^2}F_A(p^2)~,
\\[1ex]
B(p^2) &= \frac{a\pinch(p^2+m^2)}{2m^2}F_B(p^2)~,
\end{align}
\end{subequations}
where the abbreviation
\begin{equation}
  a = A(m^2) =B(m^2)
  \label{eq:aDef}
\end{equation}
was used. To provide the residue conditions (\ref{eq:A=B}) and
(\ref{eq:residue}), at the pole  both functions $F_A$ and $F_B$ must have unit
values and vanishing first derivatives. Also, they may have simple poles in the
spacelike region, at $p^2=-(2a-1)m^2$ for $F_A$ and at $p^2=-m^2$ for $F_B$,
and their combined effect must not produce an additional pole for $S$.
Determination of additional properties requires a detailed microscopic
derivation of the dressed propagator outside of the scope of the present work;
for more details of the corresponding nonlinear Dyson-Schwinger-type framework,
see Ref.~\cite{hh97}.

\subsubsection{Isolating dressing contributions}

The dressed propagator may also be written equivalently as
\begin{equation}
  S(p) = \frac{Z(p)}{\fs{p}-m}
\end{equation}
where the residual function,
\begin{equation}
  Z(p) = \frac{\fs{p}-m}{\fs{p}A(p^2)-mB(p^2)}~,
  \label{eq:Z(p)}
\end{equation}
goes to unity at the pole. Writing $Z=1+(Z-1)$, the term $Z-1$ will vanish at
the pole and thus
\begin{equation}
  S(p) = \frac{1}{\fs{p}-m} + \frac{Z(p)-1}{\fs{p}-m}
  \label{eq:poleZ}
\end{equation}
will give rise to contact-type nonpolar contribtions according to
\begin{equation}
 \frac{Z(p)-1}{\fs{p}-m}  = \frac{\fs{p}C_1(p^2) + m C_0(p^2)}{2m^2}~,
 \label{eq:Zcc}
\end{equation}
with scalar, nonpolar functions $C_0$ and $C_1$.

These functions can be easily expressed in terms of the dressing functions $A$
and $B$, however, since there is no need to do this here, we will forego a more
detailed discussion of their structure.

For later purposes, we will need finite-difference derivatives of the dressing
functions $A$ and $B$ defined as
\begin{subequations}\label{eq:DADB_FDD}
\begin{align}
  D_A(p^2) &= 2m^2 \frac{A(p^2)-a}{p^2-m^2}~,
  \\
  D_B(p^2) &= 2m^2 \frac{B(p^2)-a}{p^2-m^2}~,
  \label{eq:DADBdefined}
\end{align}
\end{subequations}
with well-defined $\frac{0}{0}$ expressions at the pole, and their difference,
\begin{align}
  D_{AB}(p^2) &= 2m^2 \frac{A(p^2)-B(p^2)}{p^2-m^2}%
  \nonumber\\
  &=D_A(p^2)-D_B(p^2)~,
  \label{eq:DABdefined}
\end{align}
which is related to the residue condition (\ref{eq:residue}) by
\begin{equation}
   D_{AB}(m^2) = 1-a~.
   \label{eq:DABresidue}
\end{equation}

\subsection{Nucleon current}

Symmetrizing the inverse propagator, the nucleon current is obtained as
\begin{align}
  J_N^\mu(p',p) &= \GDm{ \frac{ \fs{p} A(p^2) + A(p^2) \fs{p} }{2} -mB(p^2) }_D
  \nonumber\\
  &=\Gamma_N^\mu \frac{A(p'^2)+A(p^2)}{2}
  \nonumber\\
  &\quad\mbox{}
  +\frac{(p'^2+p^2) \Gamma_N^\mu +2\fs{p}' \Gamma_N^\mu \fs{p}}{4m^2} D_A(p'^2,p^2)
  \nonumber\\
  &\quad\mbox{}
  -\frac{\fs{p}' \Gamma_N^\mu+\Gamma_N^\mu \fs{p}}{2m}D_B(p'^2,p^2)~,
  \label{eq:Jnuc}
\end{align}
where the scalar dependence was evaluated according to the Dirac-particle rule
(\ref{eq:GDDirac}), employing the extended substitution rule
(\ref{eq:MSNextended}). The FDDs
\begin{equation}
D_f(p'^2,p^2) = 2m^2\frac{f(p'^2)-f(p^2)}{p'^2-p^2}~,\qtext{for} f=A,B~,
\end{equation}
here are obvious off-shell versions of Eqs.~(\ref{eq:DADB_FDD}). Hence, the
on-shell form of the current is given by
\begin{equation}
J_N^\mu(\up',\up) = Q_p \gamma^\mu + T^\mu_N(\up',\up)~,
\label{eq:JNonshellgeneric}
\end{equation}
where the underlining indicates on-shell momenta; the corresponding incoming
and outgoing spinors have been suppressed for clarity. The form-factor
functions contained in $T^\mu_N$ according to (\ref{eq:emFFexpand}) will  be
constrained by their relation to Dirac and Pauli form factors, as given in
Sec.~\ref{sec:FF}.

Pulling out off-shell factors $(\fs{p}'-m)$ and  $(\fs{p}-m)$ on the left and
right, respectively, the whole current may be written as
\begin{align}
  J_N^\mu(p',p)&= \Gamma_N^\mu(p',p) R(p'^2,p^2)
  \nonumber\\[2ex]
  &\qbox
  +\left[\frac{\fs{p}'-m }{2m} \Gamma_N^\mu(p',p) +\Gamma_N^\mu(p',p)\frac{\fs{p}-m}{2m}\right]
  \nonumber\\[1.5ex]
  &\qquad\mbox{}\times \Big[D_A(p'^2,p^2)-D_B(p'^2,p^2)\Big]
  \nonumber\\[2ex]
  &\qbox
  +2  \frac{\fs{p}'-m}{2m}\Gamma_N^\mu(p',p) \frac{\fs{p}-m}{2m} D_A(p'^2,p^2)~,
  \label{eq:Jfinal}
\end{align}
where
\begin{equation}
  R(p'^2,p^2)=\hA(p'^2,p^2)+D_A(p'^2,p^2)-D_B(p'^2,p^2)~,
\end{equation}
with
\begin{equation}
  \hA(p'^2,p^2) = \frac{(p'^2-m^2)A(p'^2)-(p^2-m^2)A(p^2)}{p'^2-p^2}~.
\end{equation}
The latter is a well-defined FDD with half-on-shell limits
\begin{equation}
\hA(p'^2,p^2) = \begin{cases}
  A(p'^2)  &\qtext{for} p^2=m^2~,
\\
  A(p^2)  &\qtext{for} p'^2=m^2~.
\end{cases}
\end{equation}
The entire function $R(p'^2,p^2)$, therefore, goes to unity on shell because of
the residue condition (\ref{eq:residue}) producing the on-shell result
(\ref{eq:JNonshellgeneric}).

Equation (\ref{eq:Jfinal}) cleanly separates on-shell, half-off-shell, and
fully off-shell contributions here in a manner that is useful for further
applications. Even though it may not be immediately obvious from its
appearance, it does satisfy the WTI~\cite{WTI} with the fully dressed
propagator of Eq.~(\ref{eq:NucProp}),
\begin{equation}
k_\mu J_N^\mu(p',p) = Q_N \left[S^{-1}(p') - S^{-1}(p)\right]~,
 \label{eq:WTIspinhalf}
\end{equation}
as mandated by local gauge invariance. Only the $Q_p\gamma^\mu $ part of the
elementary current $\Gamma^\mu$ contributes to this result, of course. Any
information about electromagnetic structure is transverse and does not enter
the WTI.

The current as written here in (\ref{eq:Jfinal}) is similar in operator
structure to the most general ansatz discussed in Ref.~\cite{Bincer}. Since it
reproduces the WTI (\ref{eq:WTIspinhalf}) for the fully dressed propagator, its
structure clearly exhausts the necessary dependence on dressing functions $A$
and $B$. The possibility of additional terms --- which would necessarily have
to be off shell and transverse --- will be discussed in the summarizing
Sec.~\ref{sec:summary}.

\subsubsection{Relationship to Ball-Chiu current}\label{sec:BCC}

If we switch off the additional transverse piece $T^\mu_N$ for the moment,
producing $J_N^\mu\to J_0^\mu$, the remaining current for the proton may be
written as
\begin{align}
  J_0^\mu(p',p) &= J^\mu_{\textsc{bc}}(p',p)
  \nonumber\\
  &\quad\mbox{}
  + Q_p\frac{\fs{p}'i\sigma^{\mu\nu}k_\nu+i\sigma^{\mu\nu}k_\nu \fs{p}}{4m^2}D_A(p'^2,p^2)
  \nonumber\\
  &\quad\mbox{}
  -Q_p\frac{i\sigma^{\mu\nu}k_\nu}{2m}D_B(p'^2,p^2)~,
  \label{eq:J0-BC}
\end{align}
where, suppressing arguments of $D_A$ and $D_B$ for simplicity,
\begin{align}
  J^\mu_{\textsc{bc}}(p',p)&=
   Q_p\gamma^\mu\frac{A(p'^2)+A(p^2)}{2}
     \nonumber\\
  &\quad\mbox{}
+Q_p \frac{(p'+p)^\mu}{2m} \Bigg[\frac{\fs{p}'+\fs{p}}{2m}D_A
  -D_B \Bigg]
  \nonumber\\
  &=Q_p (p'+p)^\mu\frac{S^{-1}(p')-S^{-1}(p)}{p'^2-p^2}
  \nonumber\\
  &\quad\mbox{}
  +Q_p\left[\gamma^\mu -\frac{(p'+p)^\mu}{p'^2-p^2}  \fs{k}\right]\frac{A(p'^2)+A(p^2)}{2}~.
\end{align}
This current was proposed by Ball and Chiu~\cite{BC80} as one of the simplest
nonsingular symmetric expressions whose four-divergence provides the WTI for
the fully dressed propagator (\ref{eq:NucProp}). The additional transverse term
in the last expression is needed to cancel the singularity of the first term at
$p'^2=p^2$.

These relations shows most clearly that all three currents ---
$J^\mu_{\textsc{bc}}$, $J^\mu_0$, and $J^\mu_N$ --- satisfy the WTI
(\ref{eq:WTIspinhalf}) because they differ from each other by manifestly
transverse contributions.

We see here that the Ball-Chiu current is obtained as
\begin{equation}
J^\mu_{\textsc{bc}}(p',p) = \GDm{S^{-1}(p)}_S~,
\end{equation}
whereas
\begin{equation}
J^\mu_0(p',p) = \GDm{S^{-1}(p)}_D~,
\end{equation}
i.e., they differ by how the electromagnetic field is being coupled to their
scalar parts. If we consider their respective on-shell forms, we find
\begin{equation}
J^\mu_{\textsc{bc}}(\up',\up) =  Q_p \gamma^\mu + Q_p \frac{i\sigma^{\mu\nu}k_\nu}{2m} (a-1)
\label{eq:BConshell}
\end{equation}
and
\begin{equation}
  J^\mu_0(\up',\up) =   Q_p \gamma^\mu~,
  \label{eq:J0onshell}
\end{equation}
where the additional transverse $\sigma^{\mu\nu}k_\nu$ contributions in
(\ref{eq:J0-BC}) that originate from the proper Dirac treatment cancel the
$(a-1)$ term in (\ref{eq:BConshell}). Clearly, Eq.~(\ref{eq:J0onshell})
provides the correct on-shell limit if one switches off all transverse pieces
in the substitution rule (\ref{eq:MSNextended}). (This result is also obtained
if $a=1$, of course, but this value is \emph{not} required by the residue
conditions.)

\subsubsection{Determining Dirac and Pauli form factors}\label{sec:FF}

Writing out the specific on-shell forms of $J^\mu_N$ for proton and neutron,
\begin{equation}
J_N^\mu(\up',\up) =
\begin{cases}\displaystyle
e\gamma^\mu (1+ f^p_1)  + e\st^\mu f^p_2
&\quad \text{(proton)}~,
\\[2ex]
\displaystyle
e\gamma^\mu f^n_1  + e\st^\mu  f^n_2
&\quad \text{(neutron)}~,
\end{cases}
\label{eq:Jw/f}
\end{equation}
and comparing this with the usual expressions,
\begin{subequations}\label{eq:Cexp2}
\begin{align}
  J^\mu_{p}(\up',\up) &= e \gamma^\mu F^p_1(k^2) +e \st^\mu \kappa_p F^p_2(k^2)~,
  \\[1ex]
  J^\mu_{n}(\up',\up) &= e \gamma^\mu F^n_1(k^2) + e \st^\mu \kappa_n  F^n_2(k^2)~,
\end{align}
\end{subequations}
where $F_1^N(k^2)$ and $F_2^N(k^2)$ are the Dirac and Pauli form factors, with
$\kappa_p$ and $\kappa_n$ being the anomalous magnetic moments of proton and
neutron, respectively, this produces the identifications
\begin{subequations}\label{eq:emFFnormalized}
\begin{align}
f^p_1(m^2,m^2;k^2)&= F^p_1(k^2) - 1~,
\\[1ex]
f^p_2(m^2,m^2;k^2) &= \kappa_p F^p_2(k^2)~,
\\[1ex]
f^n_1(m^2,m^2;k^2) &= F^n_1(k^2)~,
\\[1ex]
f^n_2(m^2,m^2;k^2) &= \kappa_n  F^n_2(k^2)~.
\end{align}
\end{subequations}
In view of the normalizations
\begin{subequations}
\begin{align}
  F^n_1(0)&=0~, & F^n_2(0)&=1~,
  \\
  F^p_1(0)&=1~, & F^p_2(0)&=1~,
\end{align}
\end{subequations}
the two Dirac coefficient functions $f^N_1$ vanish at $k^2=0$, cancelling the
$1/k^2$ singularity in $\gt^\mu$; the two Pauli coefficient functions $f^n_2$
produce the respective anomalous magnetic moment at $k^2=0$.

With the on-shell identifications (\ref{eq:emFFnormalized}), the nucleon
current (\ref{eq:Jnuc}) thus reproduces the correct experimental information by
construction.

Allowing for off-shell nucleons, the form factors $f_i^N$ ($i=1,2$) may be
written without lack of generality as
\begin{subequations}
\begin{align}
  f^N_1(p'^2,p^2;k^2) & = \frac{k^2}{m^2} H^N_1(p'^2,p^2;k^2)~,
  \\
  f^N_2(p'^2,p^2;k^2) & = 1+\frac{k^2}{m^2} H^N_2(p'^2,p^2;k^2)~,
\end{align}
\end{subequations}
for $N=p,n$, with nonsingular symmetric scalar functions $H^N_i$, for $i=1,2$.
The current operator $\Gamma^\mu_N$ of Eq.~(\ref{eq:MSNextended}) then reads in
full detail
\begin{align}
  \Gamma_N^\mu(p',p) & =\gamma^\mu\left[Q_p + Q_N\frac{k^2}{m^2}H^N_1(p'^2,p^2;k^2)\right]
  \nonumber\\
  &\quad\mbox{}
  +Q_N \kappa_N \frac{i\sigma^{\mu\nu}k_\nu}{2m} \left[1 + \frac{k^2}{m^2}H^N_2(p'^2,p^2;k^2)\right]
  \nonumber\\
  &\quad\mbox{}
   + Q_N k^\mu \frac{\fs{p}'-\fs{p}}{m^2} H^N_1(p'^2,p^2;k^2)~.
\end{align}
We emphasize here that electromagnetic structure information enters this
expression only in manifestly transverse terms. We also note that this is not
obvious if one drops the $k^\mu$ term prematurely.

\subsection{Half-on-shell contribution}\label{sec:halfnucleon}

Let us consider the half-on-shell version of the current (\ref{eq:Jfinal})
where the incoming nucleon is on shell. The kinematics then are the same as for
the $s$-channel diagram in Fig.~\ref{fig:photo}. One finds
\begin{align}
  J_N^\mu(p_s,\up)&= \left[ R(s)
  +\frac{\fs{p}_s-m }{2m}  D_{AB}(s)\right] \Gamma_N^\mu(p_s,\up)~,
  \label{eq:JfinalRightOS}
\end{align}
where $p_s=p+k$ and $s=p_s^2$ and $R(s)$ is shorthand for $R(s,m^2)$.
Multiplying this by the $s$-channel propagator $S(p_s)$ for the fully dressed
intermediate nucleon, one obtains, after tedious, but straightforward, algebra,
\begin{align}
 S(p_s) J_N^\mu(p_s,\up) &= \frac{1}{\fs{p}_s-m} \Gamma_N^\mu (p_s,\up)
 \nonumber\\
 &=\frac{Q_p}{\fs{p}_s-m}\gamma^\mu + \frac{1}{\fs{p}_s-m} T_N^\mu (p_s,\up)~.
  \label{eq:SJNhalfon}
\end{align}
All dressing effects here cancel fully. This result is exact for the
current~(\ref{eq:Jfinal}) constructed here. The only dressing effects left here
are electromagnetic in nature, in the two half-on-shell form factors
$f^N_i(s,m^2;k^2)$ ($i=1,2$) within $T_N^\mu$.

As was the case for the pion treatment in Sec.~\ref{sec:pion}, structure
information enters the half-on-shell element only via manifestly transverse
terms, and the four-divergence here indeed produces
\begin{equation}
  k_\mu\,   S(p_s) J_N^\mu(p_s,\up) =Q_p~,
  \label{eq:kSJNonshell}
\end{equation}
as demanded by local gauge invariance.

The form factors that enter the transverse currents $\Gamma_N^\mu$ are only
needed here half on shell. Using the physical limits (\ref{eq:emFFnormalized}),
we may expand the $s$-dependent off-shell side of the equation in terms of
well-defined FDDs and write the half-on-shell contribution as
\begin{equation}
S(p_s)J_N^\mu(p_s,\up) = \frac{Q_N}{\fs{p}_s-m}\Gamma^\mu_{N,0}(k)
+ \frac{k^2}{m^3}Q_N C_N^\mu(p_s;k)~,
\label{eq:SJfinal}
\end{equation}
where the polar term comprises the usual on-shell nucleon current,
\begin{equation}
  \Gamma^\mu_{N,0}(k) = \gamma^\mu F^N_1(k^2)
  + \kappa_N \frac{i\sigma^{\mu\nu}k_\nu}{2m}  F^N_2(k^2)~,
\label{eq:NphysCurrent}
\end{equation}
and the contact term reads
\begin{align}
C^\mu_N(p_s;k)
  &=  \frac{\fs{p}_s+m}{2m}\left[\gamma^\mu D_1^N(s;k^2) + \frac{i\sigma^{\mu\nu}k_\nu}{2m}  D_2^N(s;k^2) \right] ~,
\end{align}
with FDDs
\begin{equation}
  D^N_i(s;k^2) = 2 m^2 \frac{H^N_i(s;k^2)-H^N_i(m^2;k^2)}{s-m^2}~,
\label{eq:FDDemFF}
\end{equation}
for $i=1,2$, where the incoming on-shell variable ($p^2=m^2$) is suppressed.
The unphysical $k^\mu$ term was dropped in Eq.~(\ref{eq:SJfinal}), i.e., this
equation cannot be used to verify the gauge-invariance condition
(\ref{eq:kSJNonshell}).

Structurally, the half-on-shell result (\ref{eq:SJfinal}) is exactly the same
as eq.~(\ref{eq:JpionHalfOnshell2}) for the pion and --- as for the pion
--- the nucleon structure information only enters for virtual photons for the
current determined here.  Hence, any other possible structure can only come
from off-shell electromagnetic form factors $H^N_i(s;k^2)$ (for $N=p,n$;
$i=1,2$) that contribute only for electroproduction, when $k^2\neq 0$.

It should be obvious that the analogous half-on-shell cancellations can also
easily be verified for the $u$-channel process in Fig.~\ref{fig:photo}. Hence,
for all three Born-type  contributions to the pion photoproduction process,
only the respective hadronic $\pi NN$ vertices carry structure information. The
detailed dynamics of the problem, therefore, is hidden in the contact-type
current of the last diagram in Fig.~\ref{fig:photo}.

\subsection{Real Compton scattering on the nucleon}\label{sec:ComptonNucleon}

%
\begin{figure}[!t]\centering
  \includegraphics[width=.90\columnwidth,clip=]{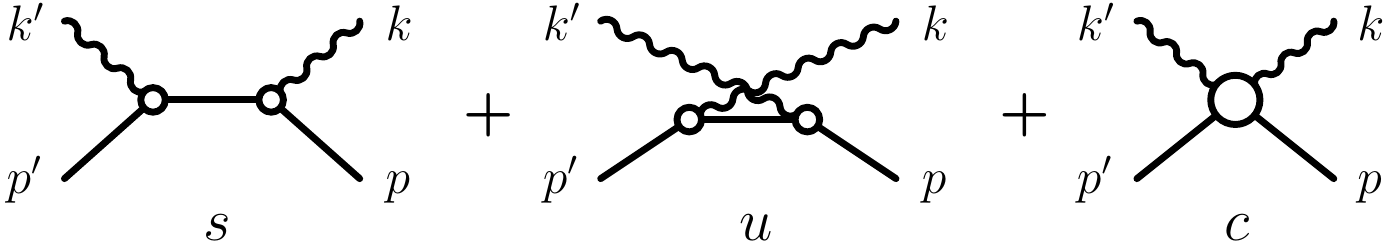}
  \caption{\label{fig:ComptonNucleon}%
  Diagrams contributing to Compton scattering on the nucleon (ignoring
  $t$-channel exchange). In
  Eqs.~(\ref{eq:ComptonNucleonSU})--(\ref{eq:ComptonNucleonAll}), the incoming
  and outgoing four-momenta are $k^\mu$ and $k'^\nu$, respectively.}
\bigskip
  \includegraphics[width=.95\columnwidth,clip=]{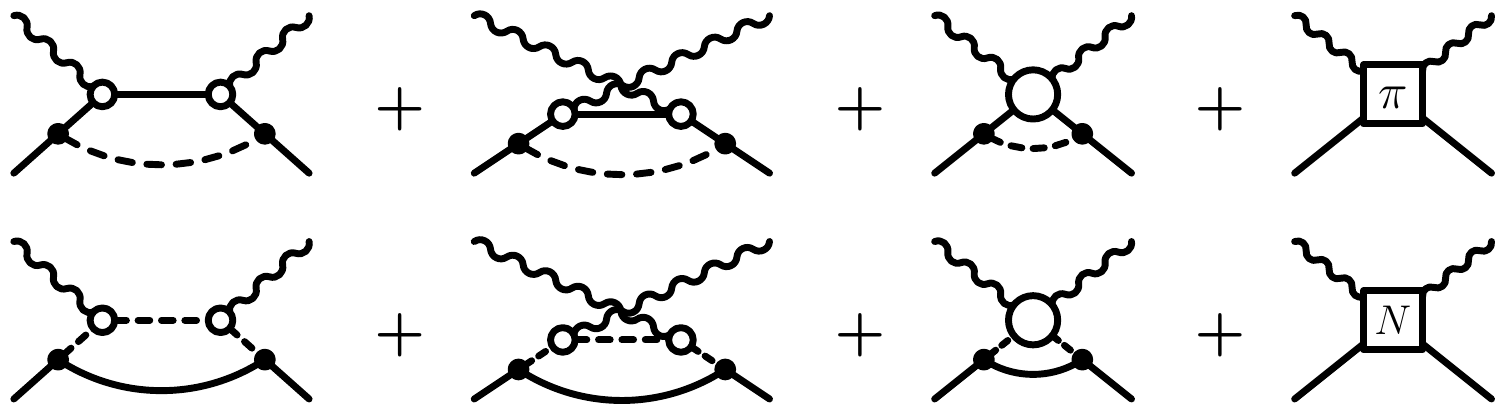}
  \caption{\label{fig:ComptonNucleonDressed}%
  Examples of genuine two-photon processes contributing to Compton scattering
  on the nucleon. The first line of diagrams provides the crossing-symmetric
  dressing of the amplitude depicted in Fig.~\ref{fig:ComptonNucleon} by a
  meson loop and the second line corresponds to intermediate Compton scattering
  on the pion as depicted in Fig.~\ref{fig:ComptonPion}. The contact-type
  squares labeled $\pi$ and $N$ subsume necessary mechanisms to
  make the contribution of each respective line transverse.}
\end{figure}
%

Let us now consider real Compton scattering on the proton using the current
(\ref{eq:Jfinal}). We may then replace $Q_N$  by $e$ and put $k^2=k'^2=0$
everywhere. Moreover, taking $k^\mu$ and $k'^\nu$ to be the four-momenta for
outgoing and incoming photons, respectively, we may use
\begin{equation}
  \Gamma^\mu_p \equiv \Gamma^\mu_{p,0}(k)
  \qtext{and}
  \Gamma^\nu_p \equiv \Gamma^\nu_{p,0}(k')
\end{equation}
as shorthand notations for the corresponding incoming and outgoing currents.
Further, denoting the intermediate four-momenta in the $s$- and $u$-channel
diagrams of Fig.~\ref{fig:ComptonNucleon} by
\begin{equation}
  p_s = p+k=p'+k'
  ~~\text{and}~~
  p_u = p'-k=p-k'~,
\end{equation}
with Mandelstam variables $s=p^2_s$ and $u=p^2_u$, and using the half-on-shell
results (\ref{eq:JfinalRightOS}) and (\ref{eq:SJfinal}), the $s$- and
$u$-channel parts of the Compton tensor read
\begin{subequations}\label{eq:ComptonNucleonSU}
\begin{align}
  M^{\nu\mu}_s(p',p) &=\Gamma^\nu_{p}\frac{R(s)}{\fs{p}_s-m}\Gamma^\mu_{p}
      +\Gamma^\nu_{p}\frac{D_{AB}(s)}{2m}\Gamma^\mu_{p}
      \nonumber\\
   &=\Gamma^\nu_{p}\frac{1}{\fs{p}_s-m}\Gamma^\mu_{p}
  \nonumber\\
  &\quad\mbox{}
      +\Gamma^\nu_{p}\left[\frac{\fs{p}_s+m}{2m^2}D_R(s)+\frac{D_{AB}(s)}{2m}\right]\Gamma^\mu_{p}~,
  \\
  M^{\nu\mu}_u(p',p) &=\Gamma^\mu_{p}\frac{R(u)}{\fs{p}_u-m}\Gamma^\nu_{p}
      +\Gamma^\mu_{p}\frac{D_{AB}(u)}{2m}\Gamma^\nu_{p}
      \nonumber\\
   &=\Gamma^\mu_{p}\frac{1}{\fs{p}_u-m}\Gamma^\nu_{p}
  \nonumber\\
  &\quad\mbox{}
      +\Gamma^\mu_{p}\left[\frac{\fs{p}_u+m}{2m^2}D_R(u)+\frac{D_{AB}(u)}{2m}\right]\Gamma^\nu_{p}~,
\end{align}
\end{subequations}
where the FDD
\begin{equation}
  D_R(x) = 2m^2\frac{R(x)-1}{x-m^2}~,\qtext{for} x=s,u~,
\end{equation}
was introduced. The respective second equalities in (\ref{eq:ComptonNucleonSU})
separate pole terms from contact terms. The contact current is obtained by the
same construction used already for the pion contact term
(\ref{eq:ComptonPionContact}). One obtains
\begin{align}
  M^{\nu\mu}_c&= -\frac{1}{2}\left[\GDn{J^\mu_N(p_\toff,\up)}+\GDm{J^\nu_N(p_\toff,\up)}\right]
\nonumber\\
&\quad\mbox{}
-\frac{1}{2}\left[\GDn{J^\mu_N(\up',p_\toff)}+\GDm{J^\nu_N(\up',p_\toff)}\right]
\nonumber\\
&= -\Gamma^\nu_{p}\left[\frac{\fs{p}_s+m}{2m^2}D_R(s)+\frac{D_{AB}(s)}{2m}\right]\Gamma^\mu_{p}
\nonumber\\
&\quad\mbox{}
-\Gamma^\mu_{p}\left[\frac{\fs{p}_u+m}{2m^2}D_R(u)+\frac{D_{AB}(u)}{2m}\right]\Gamma^\nu_{p}~,
\end{align}
which exactly cancels the contact terms in the $s$- and $u$-channel
contributions. The Compton tensor for the diagrams in
Fig.~\ref{fig:ComptonNucleon} then simply reads
\begin{align}
M^{\nu\mu}_N &= M^{\nu\mu}_s+M^{\nu\mu}_u+M^{\nu\mu}_c
\nonumber\\
&=\Gamma^\nu_{p}\frac{1}{\fs{p}_s-m}\Gamma^\mu_{p}+\Gamma^\mu_{p}\frac{1}{\fs{p}_u-m}\Gamma^\nu_{p}~,
\label{eq:ComptonNucleonAll}
\end{align}
leaving only undressed $s$- and $u$-channel terms which, taken together, are
trivially gauge invariant.

This bare expression provides the Powell cross-section~\cite{Pow49}, but it
does not describe more complex experimental data like electric and magnetic
polarizibilities (see Refs.~\cite{Lvov93,FWtext,GMP18} and references therein).
However, as with the real Compton scattering tensor for the pion discussed in
Sec.~\ref{sec:ComptonPion}, the present construction corresponds to sequential
single-photon processes. Genuine two-photon processes like those given by the
examples of dressing mechanisms depicted in
Fig.~\ref{fig:ComptonNucleonDressed}, or processes (not shown) with other
intermediate resonant  baryonic states  ($\Delta$ etc.), are not taken into
account. It is well known that such mechanisms will indeed describe
polarizabilities. Note also that the Compton tensors dressed by such loop
mechanisms are \emph{not} the undressed ones. Their external hadron lines are
off-shell, thus making them fully dependent on the propagator dressing
functions $\Pi$, for the pion, and $A$ and $B$, for the nucleon.

\section{Summary and Discussion}\label{sec:summary}

We have presented here an extension of the usual minimal substitution procedure
that provides a straightforward inclusion of electromagnetic form factors into
hadronic current operators. The important starting point here is the --- known,
but oftentimes forgotten --- fact that electromagnetic structure information
\emph{always} is limited to manifestly transverse current contributions. This
follows simply from the fact that the only electromagnetic structure
information that enters the Ward-Takahashi identities for electromagnetic
currents are the respective charges --- electromagnetic form factors do not
enter [cf.\ Eq.~(\ref{eq:WTIgeneric})].

The resulting extended substitution ans\"atze proposed in
Sec.~\ref{sec:extsub}, therefore, concern only manifestly transverse additions
to the respective basic currents for spin-0 and and spin-1/2 particles that
result from Feynman rules. Phenomenological additions of electromagnetic form
factors are nothing new and have been undertaken before. The novel aspect of
the present approach is the consequent application of this extended
electromagnetic substitution with dressed hadronic propagators, providing
currents that incorporate all dressing effects in a consistent manner.

The gauge-derivative procedure of Ref.~\cite{hh97} used in determining the
consistent currents also means, as shown in Sec.~\ref{sec:extsub}, that
coupling the electromagnetic field to scalar dressing contributions needs to be
treated differently for spin-0 and spin-1/2 particles, lest one ignores the
fact that spin-1/2 solutions also must solve the Klein-Gordon equation. As a
consequence, we showed in Sec.~\ref{sec:BCC} that the Ball-Chiu
current~\cite{BC80} lacks transverse contributions that, when added, provide
the correct on-shell limit.

Regarding the consequences of dressing, we emphasize that, although we started
out by assuming that particle propagators are fully dressed, the current
expressions obtained by taking their gauge derivatives do \emph{not} make any
assumptions about how the corresponding dressing functions are obtained. The
only features that matter are the respective residue conditions. Therefore,
\emph{any} determination of hadronic dressing effects that meets these
conditions, whether by simple single-loop models or sophisticated self-energy
expansions to all orders, can be accommodated, thus making the cancelation
effects between dressed propagators and correspondingly determined currents
found here true for any of such dressings. The cancelations are also
independent of the detailed extended substitution features. In other words,
they are also true if one simply applies minimal substitutions, without any
extensions, to the respective dressed propagators.

Note that the procedure for the current construction used here, while always
producing a fully locally gauge-invariant current, in general does not preclude
the possibility of additional \emph{transverse off-shell} contributions, where
transversality and off-shellness are both essential requirements. Such currents
would spoil the perfect cancelations, of course, but they cannot be excluded on
general grounds. However, as discussed in Refs.~\cite{Lvov93,SF95,Fea98,FS00},
\emph{any}  reaction-dynamical description is subject to
representation-dependent ambiguities because one can \emph{always} trade off
the off-shell dependence of pole terms against associated contact-type
contributions by using appropriate finite-difference derivatives. The
ubiquitous use of FDDs in the present formulation is testament to this
ambiguity.

In any formulation of the reaction dynamics of a photoprocess, the question,
therefore, is \emph{not} whether a particular off-shell or transverse current
has been included for the photoprocess at hand, but whether all effects
--- whether of the polar-type or contact-type --- have been included
\emph{consistently}. In this respect, therefore, their is no need to consider
additional transverse contributions for the currents derived here consistently
with their fully dressed hadron propagators because any additional contribution
deemed necessary for the description of the physics at hand can be expressed in
terms of contact terms. The hadron loops around the (off-shell) Compton tensors
in Figs.~\ref{fig:ComptonPionDressed} and \ref{fig:ComptonNucleonDressed}
provide examples of this kind.

In view of the obvious representation dependence of off-shell polar
\textit{vs.} contact-term contributions discussed above, perhaps the most
surprising aspect of the present findings is that --- for real photons at least
--- the dressing cancellations found here \emph{force} a natural split into pure pole
terms and residual contact-type contributions in certain dynamical on-shell
situation, even if one starts out assuming fully dressed hadrons. The key to
this result is the consistency of the currents derived in the present framework
with whatever dressing goes into the hadron propagators.

We add here that for processes like pion production off the nucleon, as
depicted in Fig.~\ref{fig:photo}, the dressing cancellations found here mean
that for the usual $s$-, $u$-, and $t$-channel terms, the only relevant
dressing contributions are those stemming from the hadronic $\pi NN$ vertex.
This may at least partially explain the relative success of phenomenological
approaches to photo- and electroproduction of mesons that model the hadronic
vertex by a simple cutoff function and completely ignore any other dressing
effects. Furthermore, as a consequence of the present findings, this means
that, other than adding baryonic resonance content, most of the effort in
describing the detailed dynamics of meson-production reactions needs to go into
the determination of \emph{transverse} contact-type currents that arise from
final-state interactions. This is indeed the approach advocated and executed in
Refs.~\cite{hh00,HNK06,HHN11,HDH11}, where it was based on gauge-invariance
considerations alone, without the benefit of the present insights.

Finally, we have not yet considered electromagnetic processes involving
particles with higher spins beyond 1/2 along the lines presented here, but we
expect this should be possible as well.

\acknowledgments The author gratefully acknowledges discussions with Kanzo
Nakayama and Harald Griesshammer. The present work was partially supported by
the U.S. Department of Energy, Office of Science, Office of Nuclear Physics,
under Award No.\ DE-SC0016582.

\appendix
\section{Gauge derivative and minimal substitution}\label{app:MS}

To make the present paper self-contained, we recapitulate here some basic
features of the gauge-derivative device introducing in Ref.~\cite{hh97}, where
full details can be found. The gauge derivative provides a convenient shorthand
procedure for implementing minimal substitution in settings where the external
electromagnetic field $A^\mu$ interacts with larger systems of strongly
interacting particles. The properties of such systems are fully described by
their connected Green's functions. As described in the Introduction, the
coupling of the electromagnetic field is usually effected by an LSZ-type
reduction of the gauged Green's function~\cite{Wtext,PStext,LSZ}. However, to
maintain full local gauge invariance, in principle, this should be done at
every level of all details that enter the microscopic description of the
reaction at hand. The gauge derivative is designed to make bookkeeping easier,
by --- loosely speaking --- allowing attaching a photon to every particle with
charge $Q$ and injecting the photon's four-momentum such that it is available
to every reaction mechanism `downstream' of the initial photon interaction.

The basic definition of the gauge-derivative braces $\GDm{\cdots}$ is given by
the functional derivative of the minimal-substitution rule (\ref{eq:MSdefined})
for a particle of four-momentum $p$ with charge operator $Q$,
\begin{equation}
  \GDm{p^\nu}=-\frac{\delta}{\delta A_\mu} \left(p^\nu - Q A^\nu\right) = Q g^{\mu\nu}~.
  \label{eq:GDBasicDefinition}
\end{equation}
Since it is a derivative, the product rule applies providing
\begin{align}
  \GDm{p^2} = g_{\lambda\nu} \GDm{p^\lambda p^\nu}
  &= g_{\lambda\nu}\left[\GDm{p^\lambda}p^\nu+p'^\lambda\GDm{p^\nu}\right]
  \nonumber\\
  &= Q(p'+p)^\mu~,
  \label{eq:GDbasicscalar}
\end{align}
where the four-momentum downstream of where the gauge derivative is applied is
increased by the photon's four-momentum $k=p'-p$. Moreover, since the gauge
derivative acts only on momenta, one has
\begin{equation}
  \GDm{\fs{p}} = \gamma_\nu\GDm{p^\nu} = Q \gamma^\mu~.
  \label{eq:GDbasicDirac}
\end{equation}
These two results provide the basic coupling mechanisms for scalar and Dirac
particles, respectively, given in Eqs.~(\ref{eq:MSbasicPion}) and
(\ref{eq:MSbasicNucleon}) of Sec.~\ref{sec:extsub}.

The product rule also applies to any functions $f(p)$ and $g(p)$ of the
four-momentum, i.e.,
\begin{equation}
  \GDm{f(p)g(p)} = \GDm{f(p)} g(p) +f(p')\GDm{g(p)}~,
  \label{eq:GDbasicproduct}
\end{equation}
and if the functions commute, symmetrization is required to prevent
ambiguities,
\begin{align}
\GDm{f(p)g(p)} \to &\GDm{\frac{f(p)g(p)+g(p)f(p)}{2}}
\nonumber\\
   &=  \GDm{f(p)}\frac{g(p')+g(p)}{2}
\nonumber\\
   &\qquad\mbox{}+  \GDm{g(p)}\frac{f(p')+f(p)}{2}~.
\end{align}
Generally, unsymmetrized results differ by transverse terms from symmetrized
ones.

Applying the product rule to
\begin{equation}
\GDm{t(p)\,t^{-1}(p)} =\GDm{1}=0~,
\end{equation}
where $t(p)$ is a generic propagator for a particle with four-momentum $p$, one
immediately finds that the electromagnetic current for this particle is
determined by the generic expression
\begin{equation}
J^\mu =  -t^{-1}(p')\GDm{t(p)} t^{-1}(p) =  \GDm{t^{-1}(p)}~,
\label{eq:GDpropagator}
\end{equation}
where the current definition follows from the LSZ procedure~\cite{LSZ} since
$t(p)$ is the two-point Green's function for single-particle propagation. For
scalar and Dirac particles, this respectively provides
Eqs.~(\ref{eq:GDbasicscalar} ) and (\ref{eq:GDbasicDirac}) as the basic
coupling mechanisms since their inverse undressed propagators are given by
$(p^2-m^2)$ and $(\fs{p}-m)$, respectively. The generic Ward-Takahashi
identity~\cite{WTI} for the current (\ref{eq:GDpropagator}) is given by
\begin{equation}
  k_\mu J^\mu = Q\left[t^{-1}(p')-t^{-1}(p)\right]~,
  \label{eq:WTIgeneric}
\end{equation}
which in addition to the static charge $Q$ only retains the hadronic dressing
information that resides in the propagator. No other electromagnetic
information enters here. The WTI (\ref{eq:WTIgeneric}) is the necessary and
sufficient condition for the current $J^\mu$ to be \emph{locally} gauge
invariant. Global gauge invariance --- i.e., current conservation --- follows
trivially in the on-shell limit.

Moreover, as explained in Sec.~\ref{sec:extsub} in the context of
Eq.~(\ref{eq:DiracKG}), it is crucially important that this equality extends to
the respective current expressions. This requires that the corresponding gauge
derivatives of both sides of Eq.~(\ref{eq:DiracKG}) must be identical, leading
to the condition
\begin{align}
\GDm{\frac{1}{\fs{p}-m}} &
\stackrel{!}{=}
\frac{1}{2}\bigg\{\frac{1}{p^2-m^2}(\fs{p}+m)
\nonumber\\[1ex]
&\qquad\qquad\mbox{}+(\fs{p}+m)\frac{1}{p^2-m^2}\bigg\}^\mu
~,
\label{eq:demandequality}
\end{align}
where the right-hand side was symmetrized. The exclamation mark on top of the
equal sign indicates that this is a necessary identity. Straightforward algebra
shows then that with (\ref{eq:GDbasicDirac}) given, the coupling associated
with $p^2$ on the right-hand side must be evaluated according to the Dirac
particle rule
\begin{equation}
  \GDm{p^2}_D \equiv \GDm{\fs{p}^2}= Q\left(\fs{p}'\gamma^\mu+\gamma^\mu \fs{p}\right)~,
\end{equation}
as stated in Eq.~(\ref{eq:GDDirac}).


\end{document}